\documentclass[useAMS,usenatbib]{mn2e}

\usepackage{lscape}
\usepackage{aas_macros}
\usepackage{times}

\newbox\grsign \setbox\grsign=\hbox{$>$} \newdimen\grdimen \grdimen=\ht\grsign
\newbox\simlessbox \newbox\simgreatbox \newbox\simpropbox \newbox\wtildebox 
\setbox\simgreatbox=\hbox{\raise.5ex\hbox{$>$}\llap
    {\lower.5ex\hbox{$\sim$}}}\ht1=\grdimen\dp1=0pt
\setbox\simlessbox=\hbox{\raise.5ex\hbox{$<$}\llap
    {\lower.5ex\hbox{$\sim$}}}\ht2=\grdimen\dp2=0pt

\newcommand{\be}{\mbox{\begin{equation}}}
\newcommand{\ee}{\mbox{\end{equation}}}

\newcommand{\Cref}{\mbox{$m_{\rm ref}$}}

\newcommand{\msun}{\mbox{M$_\odot$}}

\title{Did massive black holes in globular clusters initially satisfy galactic scaling relations?}
\author{J.~M.~Diederik Kruijssen$^1$ and Nora L\"{u}tzgendorf$^2$\\
$^{1}$Max-Planck Institut f\"{u}r Astrophysik, Karl-Schwarzschild-Stra\ss e 1, 85748 Garching, Germany; kruijssen@mpa-garching.mpg.de\\
$^{2}$European Southern Observatory, Karl-Schwarzschild-Stra\ss e 2, 85748, Garching, Germany; nluetzge@eso.org}

\begin{document}

\date{Accepted 2013 May 27.  Received 2013 May 8; in original form 2013 March 25}

\pagerange{\pageref{firstpage}--\pageref{lastpage}} \pubyear{2012}
\label{firstpage}

\maketitle

\begin{abstract}
The masses of supermassive black holes (SMBHs, $M_{\rm BH}=10^6$--$10^{11}~\msun$) in the centres of galaxies are related to the host stellar spheroid mass and velocity dispersion. A key question is how these relations originate, and over which range of black hole masses they hold. It has been speculated that intermediate-mass black holes (IMBHs, $M_{\rm BH}=10^2$--$10^5~\msun$) could play a fundamental role in the growth of SMBHs. A handful of IMBHs has recently been detected in Galactic globular clusters (GCs), but their masses are inconsistent with the galactic scaling relations of SMBHs. In this Letter, we derive the initial properties of the GCs using a standard analytical evolutionary model, of which the free parameters are fixed by independent constraints. We find that the observed IMBH masses initially followed the galactic SMBH scaling relations, and subsequently moved off these relations due to the dynamical evolution of their host GCs. This work is concluded with a brief discussion of the uncertainties and the implications of our results for the possible universality of massive black hole growth.
\end{abstract}


\section{Introduction} \label{sec:intro}
The supermassive black holes (SMBHs) that reside in the centres of galaxies have masses $M_{\rm BH}$ that are tightly correlated with the properties of their host stellar spheroid, such as its velocity dispersion $\sigma$ \citep[i.e.~the $M_{\rm BH}$--$\sigma$ relation, see e.g.][]{ferrarese00,gebhardt00,tremaine02,gueltekin09} and its luminosity or stellar mass $M_{\rm sph}$ \citep[i.e.~the $M_{\rm BH}$--$M_{\rm sph}$ relation, see e.g.][]{kormendy95,magorrian98,haering04,gueltekin09}. The origin of these scaling relations is still an open question, and the answer presumably requires an understanding of the formation and growth mechanisms for SMBHs, which is thought to be driven by gas accretion and/or black hole mergers \citep[e.g.][]{silk98,dimatteo05,hopkins06,fanidakis11}. While a complete understanding of the growth rate has not yet been achieved, the presence of $\sim10^{9}$--$10^{10}~\msun$ SMBHs at $z\sim7$ \citep[e.g.][]{mortlock11} suggests that some substantial part of the SMBH growth must already have taken place in the early Universe.

In an effort to extend the $M_{\rm BH}$ scaling relations to lower masses, and to find possible building blocks of SMBHs, there has been a surge of observational work looking for intermediate-mass black holes (IMBHs) in globular clusters (GCs) \citep[e.g][]{baumgardt03a,gebhardt05,vandenbosch06,noyola08,vandermarel10,strader12,luetzgendorf12c}. The existence of IMBHs \citep{silk75} has been predicted as the result of runaway merging in dense stellar systems \citep{portegieszwart02,portegieszwart04,gurkan04}, or as the possible end products of population~III stars \citep{madau01,trenti07}. The search for IMBHs has been guided by the existing SMBH scaling relations, which predict $M_{\rm BH}=10^3$--$10^4~\msun$ for the (initial) mass range of massive GCs ($10^6$--$10^7$~\msun). At present, a handful of IMBHs have been identified through integral-field spectroscopy of GCs (see \citealt{luetzgendorf13} for a compilation), typically at the $3\sigma$ level. Other methods are generally less successful \citep[although see e.g.][]{miller03} or even present evidence against the existence of IMBHs. \citet{strader12} do not find any radio emission in the GCs M15, M19, and M22. This could be due to the well-known paucity of gas in GCs, or the possibly episodic nature of the accretion onto an IMBH -- these and other factors obstruct the straightforward detection of IMBHs in GCs \citep{miller02}.

Under the assumption that the kinematic detections of IMBHs are real, the sample of detected IMBHs has grown sufficiently to enable a statistical comparison of their masses and host GC properties to the SMBH scaling relations. The key result of this comparison is that the IMBHs have masses that lie above the SMBH scaling relations, and follow their own, shallower relations that intersect the SMBH scaling relations at $M_{\rm BH}=10^5$--$10^6~\msun$ \citep{luetzgendorf13}. However, it is known (1) that GCs must have lost a substantial fraction of their initial mass by dynamical evolution, and (2) that they must have expanded since their birth \citep{elmegreen97,vesperini97,fall01,elmegreen10,gieles11b,kruijssen12c}. Both effects imply that their position relative to the SMBH scaling relations is continuously evolving {in a way that is different from galaxies}.

In this Letter, we account for the evolution of GCs using a simple, standard evolutionary model that has been used in the past to explain several other characteristics of GCs. We find that the observed IMBH masses initially followed the galactic SMBH scaling relations, and subsequently moved off these relations due to the dynamical evolution of their host GCs -- provided that most of the IMBH growth occurred early on.

\section{Massive black hole scaling relations}

\subsection{Data sample}
We use two different samples of dynamical black-hole mass measurements: SMBHs in galaxies and IMBHs in GCs.  For the galaxies we adopt the sample summarized in \citet{mcconnell11} and \citet{mcconnell13}, including their tabulated host galaxy properties and best-fit scaling relations. The dataset comprises 72 black-hole mass and velocity dispersion measurements in massive galaxies, including new measurements as well as literature values. A subset of 35 galaxies has known bulge masses. The SMBH masses are dynamical, i.e.~they are derived by measuring the velocity dispersion profile of gas, stars, and/or masers, and by modelling the mass distribution of the galaxy. A strong rise of the velocity dispersion in the centre of the galaxy requires a high black-hole mass in order to explain the kinematic data. 

The second dataset incorporates 14 measurements and upper limits of IMBH masses in GCs \citep{noyola10,luetzgendorf11,luetzgendorf12c,luetzgendorf12}. In \citet{luetzgendorf13}, we collected the kinematic measurements of IMBH masses in GCs in order to compare their scaling relations with those of SMBHs in galaxies. The measurements were mainly taken with integral-field units (IFU) and the IMBH masses have been obtained with a similar method as is common practice for estimating galactic SMBH masses. From the IFU data, a velocity dispersion profile was obtained that was fitted with dynamical models, varying the central IMBH mass and the mass-to-light ratio. Furthermore, Monte Carlo and $N$-body simulations were used in order to exclude other possible explanations for the central rise of the velocity dispersion (e.g.~radial anisotropy), and handle the largest uncertainties such as shot noise. While the analysis has been performed using the best available data and methods, we note that the IMBH mass measurements are still under debate due to the indirect nature of their detection and contradictory results from different groups \citep[e.g.][]{hurley07,strader12}.

\subsection{Globular cluster evolution model} \label{sec:model}
We account for the dynamical evolution of GCs by comparing the observational data to cluster evolution models. We produce cluster isochrones in the $M_{\rm BH}$--$M_{\rm sph}$ plane using the semi-analytic cluster model {\sc space} \citep{kruijssen08,kruijssen09c}, which for a given, static tidal field provides a description for the mass-loss history of clusters that resembles the results of $N$-body simulations within a few per cent \citep{lamers05,lamers10}. It includes stellar evolution {\citep[adopting a metallicity of $Z=0.001$]{marigo08}}, the production and retention of stellar remnants {(with retention fractions as in \citealt{kruijssen09c})}, and the escape of stars due to two-body relaxation. {Note that the details of the stellar evolution and remnant retention model do not affect the results of this work.} The mass loss rate {due to dynamical evolution} is parametrized in the models as
\begin{equation}
\label{eq:dmdt}
\frac{{\rm d}M}{{\rm d}t}=-\frac{M}{t_{\rm dis}}=-\frac{M^{1-\gamma}}{t_0} ,
\end{equation}
where $t_{\rm dis}=t_0M^\gamma$ is the disruption time-scale, with $M$ the cluster mass in solar masses, $\gamma$ a constant in the range 0.6--1 \citep{spitzer87,lamers05}, and $t_0$ a normalization constant that represents the lifetime of a hypothetical $1~\msun$ cluster and is set by the tidal field (or angular velocity $\Omega$) as $t_0\propto\Omega^{-1}$. We note that the results presented in this work do not strongly depend on the adopted value of $\gamma$. We use $\gamma=0.7$, which implies $t_0=10.7$~Myr for the Galactic tidal field at the galactocentric radius of the solar neighbourhood \citep{kruijssen09}, {and is consistent with theory \citep{baumgardt01} and $N$-body simulations \citep{lamers10}. As shown by \citet{lamers05}, $\gamma\sim0.7$ leads to a near-linear decrease of the cluster mass with time. They also present an analytical fit to the resulting mass evolution, which agrees with the numerical integration of equation~(\ref{eq:dmdt}) to within a few per cent.}

The comparison of the GCs and their IMBHs to the galactic SMBH scaling relations only depends on the typical total mass loss of the GCs, which is set by $t_0$. Previous studies have shown that $t_0\sim1~{\rm Myr}$ (implying $t_{\rm dis}=16$~Gyr for a cluster with an initial mass of $M_{\rm init}=10^6~\msun$) is needed to describe several properties of the Galactic GC system, such as the GC mass function \citep{kruijssen09b} and their stellar composition \citep{goudfrooij13}. Such a disruption time-scale is a factor of 5--10 shorter than expected for the {\it current} orbital characteristics of the Galactic GC system. {This substantial difference should be expected, because basing the disruption time-scale on the current orbits assumes a steady state and does not account for the total amount of disruption over cosmic time.} The high-redshift, natal environment of GCs was rich in dense gas clouds, and hence much more disruptive than the smooth Galactic halo \citep{gieles06,kruijssen11}, causing much of the GC disruption to have occurred at early times \citep{elmegreen10,kruijssen12c}. We adopt the standard value of $t_0=1~{\rm Myr}$ and discuss the effect of different disruption time-scales below. {As shown in \citet{kruijssen09b}, the GC mass distribution is modelled with a similar accuracy when adopting the median disruption time-scale compared to modelling a full spectrum of disruption time-scales.} The present age of the GC population is taken to be $\tau=12~{\rm Gyr}$.

\begin{figure*}
\center\resizebox{\hsize}{!}{\includegraphics{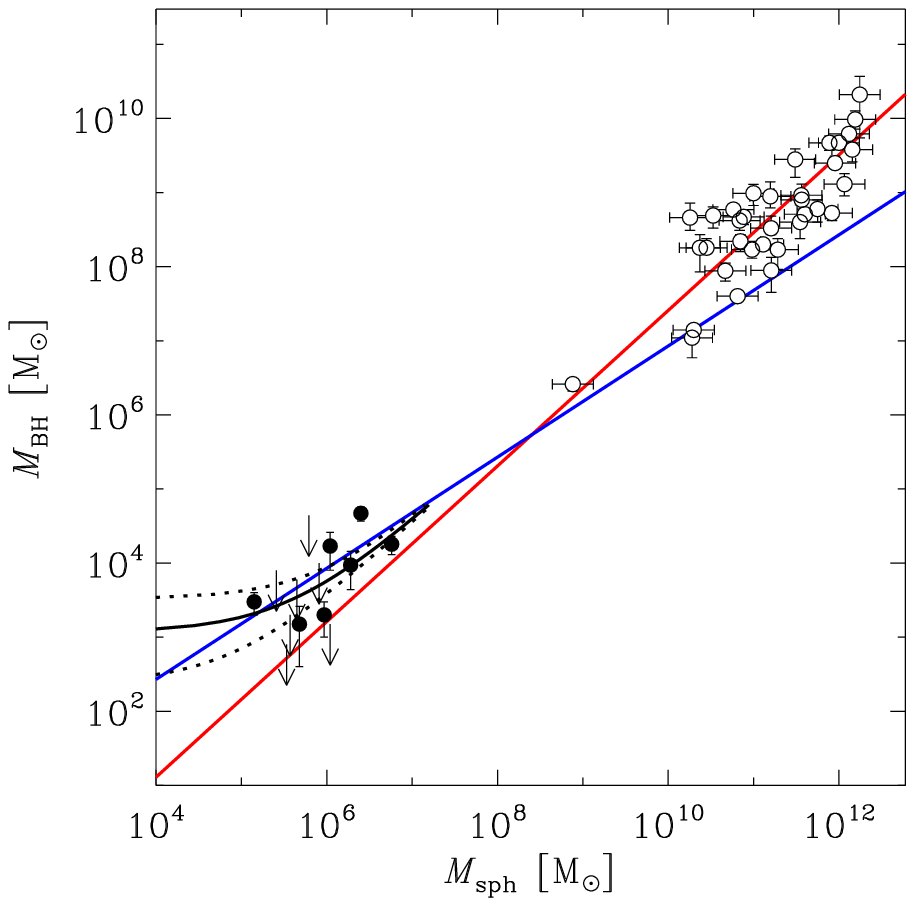}\includegraphics{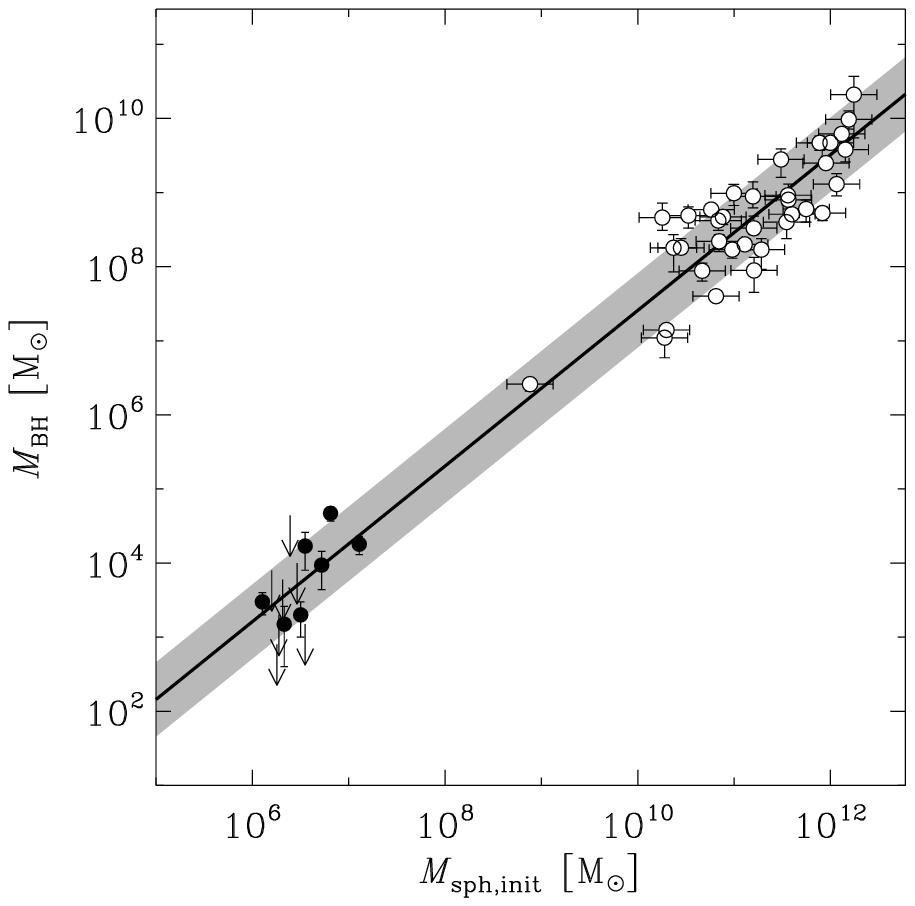}}\\
\caption[]{\label{fig:mbhmhost}
      {\it Left}: relation between black hole mass $M_{\rm BH}$ and the host spheroid mass $M_{\rm sph}$. Open circles indicate the SMBH sample from \citet{mcconnell13}, whereas filled circles represent the IMBH sample from \citet{luetzgendorf13}, with ({the tail ends of the}) arrows denoting upper limits. The red (SMBH, slope of $\beta=1.05$) and blue (IMBH, slope of $\beta=0.75$) lines are power law fits to the data ({the blue line takes into account the upper limits}). The black curves indicate the present-day relation between $M_{\rm BH}$ and $M_{\rm sph}$ if the IMBH mass is set by the {\it initial} GC mass according to the galactic (red) scaling relation. The dotted lines indicate the typical variation of $t_0$ \citep{kruijssen09b}, with $t_0=0.5~{\rm Myr}$ (top) and $t_0=3~{\rm Myr}$ (bottom). The solid line represents the standard value of $t_0=1~{\rm Myr}$ and is not a fit to the data. {\it Right}: relation between the black hole mass and the initial host spheroid mass $M_{\rm sph,init}$, which for GCs is evolved backwards in time from the present-day GC properties using $t_0=1$~Myr (solid black line in the left-hand panel). The galactic spheroid masses are unchanged, and the grey area indicates a scatter of 0.5~dex.
                 }
\end{figure*}

\subsection{A single scaling relation spanning seven decades in mass} \label{sec:msph}
We can test whether IMBHs in GCs initially followed the galactic $M_{\rm BH}$--$M_{\rm sph}$ relation by first assuming that they did, modelling the GC mass evolution due to disruption, and then comparing the resulting relation to the observed one. As detailed in \S\ref{sec:model}, we adopt a cluster evolution model in which the only important free parameter ($t_0=1~{\rm Myr}$) is set by independent observations of the GC mass function and their stellar content. This model accounts for the dynamical mass loss of the GCs, while the IMBH mass remains constant after being set by the galactic SMBH scaling relation. {Note that dynamically, it is desirable to compare present galaxies to young GCs -- because the dynamical time-scale of GCs is much shorter, a $\sim10^8$~yr-old GC and a present-day massive galaxy have similar dynamical ages of $\sim10^2$ dynamical times. Hence, the evolutionary state of the GCs should be `rewinded' before comparing them to the galactic scaling relations.}

The modelled GC isochrones in the $M_{\rm BH}$--$M_{\rm sph}$ plane are shown in the left-hand panel of Figure~\ref{fig:mbhmhost}, together with the observed black hole masses. The IMBH masses are situated above the SMBH scaling relation (red line) and follow their own, shallower relation (blue line). However, the evolved form of the galactic $M_{\rm BH}$--$M_{\rm sph}$ relation (black, solid line) agrees remarkably well with the observed GC and IMBH masses. The dotted lines indicate how the isochrones change for the expected variation of $t_0$, which is roughly consistent with the observed scatter.

The above approach can also be reversed, to show the black hole mass as a function of the {\it initial} GC mass (right-hand panel of Figure~\ref{fig:mbhmhost}). We use the black, solid line from the left-hand panel to estimate the initial masses of the observed GCs, and leave the galactic bulge masses unchanged. The figure shows that the initial GC masses are consistent with their IMBHs following the galactic $M_{\rm BH}$--$M_{\rm sph}$ relation. We quantify the agreement with a simple Kolmogorov-Smirnov (KS) test, {excluding the GCs which only have upper limits on $M_{\rm BH}$}. The respective distributions of the initial GC masses and bulge masses {\it around} the relation, i.e.~$\Delta_M\equiv \log{(M_{\rm sph,init}/M_{\rm sph,rel})}$, where $M_{\rm sph,rel}$ is the the host spheroid mass implied by the $M_{\rm BH}$--$M_{\rm sph}$ relation for a given observed black hole mass, are statistically consistent at the $p=0.88$ level, as opposed to $p=0.04$ when using the present-day GC masses. We conclude that when accounting for the dynamical evolution of GCs, the $M_{\rm BH}$--$M_{\rm sph}$ relation extends to the GC regime.
\begin{figure*}
\center\resizebox{\hsize}{!}{\includegraphics{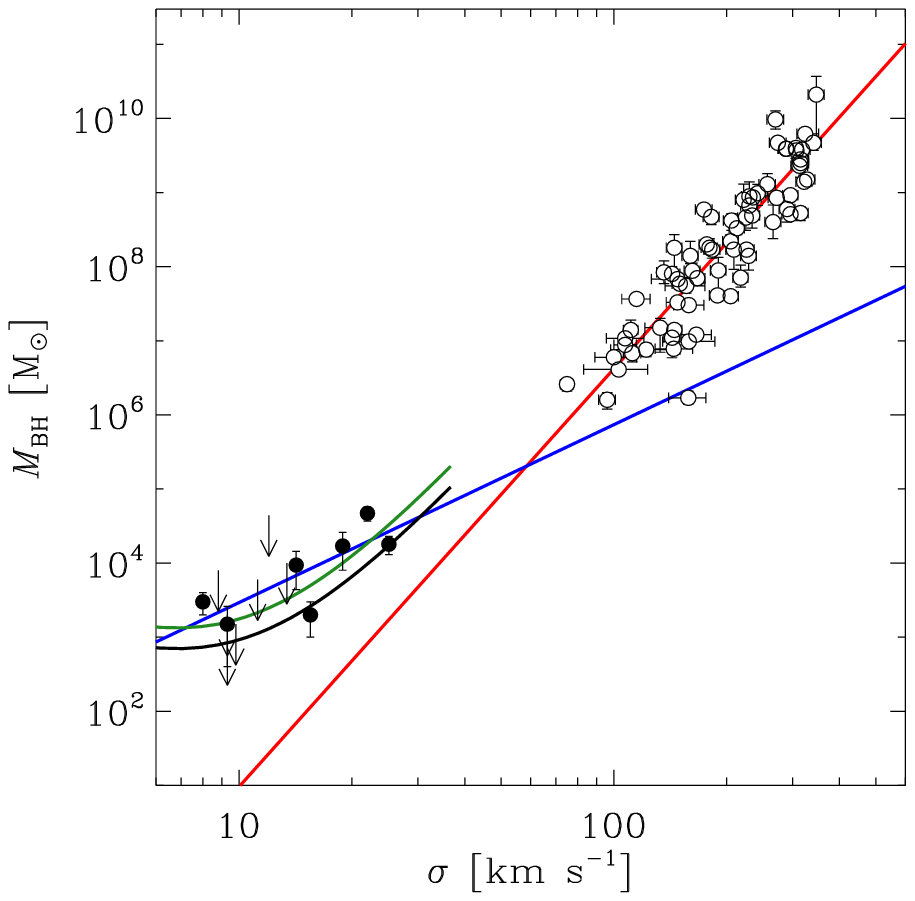}\includegraphics{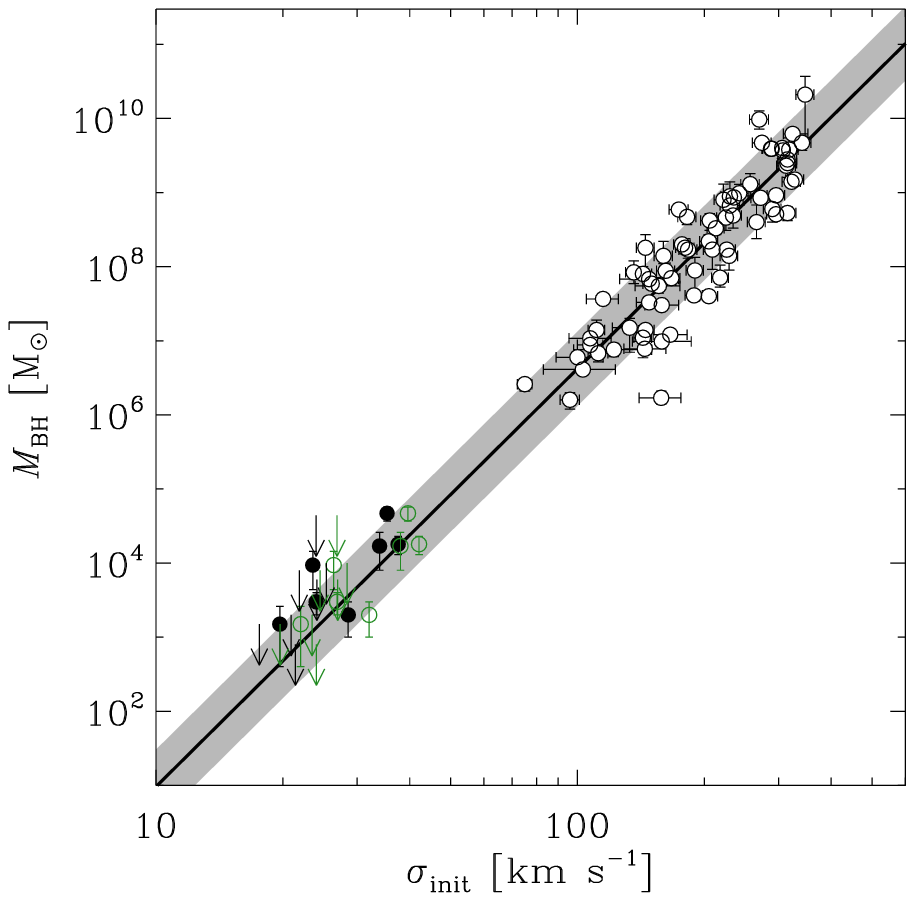}}\\
\caption[]{\label{fig:mbhsigma}
      {\it Left}: relation between black hole mass $M_{\rm BH}$ and the host spheroid velocity dispersion $\sigma$. Symbols and lines indicate the same as in Figure~\ref{fig:mbhmhost} (with slopes of $\beta=5.64$ and $\beta=2.40$ for the SMBH and IMBH relations, respectively). The black curve accounts for the GC mass loss only, whereas the green curve also includes the expansion of the GCs. Both are not fits to the data. {\it Right}: relation between the black hole mass and the {\it initial} host spheroid velocity dispersion $\sigma_{\rm init}$, which for the Galactic GCs (solid dots and arrows) is evolved backwards in time from the present-day GC properties using $t_0=1$~Myr (black line in the left-hand panel). The green symbols are similar, but also account for the GC expansion by using the green line in the left-hand panel. The galactic velocity dispersions are unchanged, and the grey area indicates a scatter of 0.5~dex.
                 }
\end{figure*}

\subsection{The black hole mass -- velocity dispersion relation} \label{sec:sigma}
Analogously to the $M_{\rm BH}$--$M_{\rm sph}$ relation in \S\ref{sec:msph}, {we can test whether the deviation of IMBHs in GCs from the galactic $M_{\rm BH}$--$\sigma$ relation is consistent with the dynamical evolution of GCs. This requires the conversion of the present-day GC velocity dispersion to its value at the time of IMBH formation. The GC sample of \citet{luetzgendorf13} follows a \citet{faber76} type of relation with $M\propto \sigma^{3.0}$, which we adopt to map the present-day model GC masses to velocity dispersions. If we make the reasonable assumption that the GCs retain virial equilibrium throughout their disruption history, the velocity dispersions at two different times $t_1$ and $t_2$ can be related:
\begin{equation}
\label{eq:sigmai}
f_\sigma(t_1,t_2)\equiv\frac{\sigma(t_2)}{\sigma(t_1)}=\left[\frac{M(t_2)}{M(t_1)}\right]^{1/2}\left[\frac{R(t_2)}{R(t_1)}\right]^{-1/2} ,
\end{equation}
where the new variable $R$ is the cluster radius. While the ratio $M(t_2)/M(t_1)$ is given by the cluster evolution models from \S\ref{sec:model}, accounting for the radius evolution requires additional, highly uncertain physics. The radius evolution is very sensitive to the cluster's detailed tidal history \citep{gieles11b}. Considering the many unknowns in the accretion history of Galactic GCs and the smaller galaxies they may have orbited in the past, this cannot be accurately constrained. We therefore focus on the cluster mass evolution model of \S\ref{sec:model}, but using the model of \citet{gieles11b}, we also illustrate the possible effect of the cluster radius evolution, with an expansion factor of $R(t_2)/R(t_1)\sim1.3$ (see Appendix~\ref{sec:radius}).} As in \S\ref{sec:msph}, any free parameters are set by independent constraints.

The modelled GC isochrones in the $M_{\rm BH}$--$\sigma$ plane are shown in the left-hand panel of Figure~\ref{fig:mbhsigma}, together with the observed black hole masses. As in Figure~\ref{fig:mbhmhost}, the IMBH masses are situated above the SMBH scaling relation (red line) and follow their own, shallower relation (blue line). However, the evolved form of the galactic $M_{\rm BH}$--$\sigma$ relation including cluster mass loss only (black line) agrees remarkably well with the observed GC and IMBH masses. The green line shows the relation when a rough estimate of the typical GC expansion since their accretion into the Galactic halo is also included, and gives somewhat better agreement (see below).

The black hole mass is shown as a function of the {\it initial} GC velocity dispersion $\sigma_{\rm init}$ in the right-hand panel of Figure~\ref{fig:mbhsigma}, using the black and green lines from the left-hand panel to derive $\sigma_{\rm init}$ (black and green symbols, respectively). As before, the galactic velocity dispersions are left unchanged. The figure shows that the initial GC velocity dispersions are consistent with their IMBHs following the galactic $M_{\rm BH}$--$\sigma$ relation, depending only weakly on whether or not their expansion is included. We again run KS tests of the differences between the power law relation and the respective distributions of the initial GC and galactic velocity dispersions, i.e.~$\Delta_\sigma\equiv \log{(\sigma_{\rm init}/\sigma_{\rm rel})}$, {excluding the GCs which only have upper limits on $M_{\rm BH}$}. These clearly show that the GCs are statistically consistent with the galactic relation -- accounting for the mass evolution only gives $p=0.09$, whereas also including some GC expansion yields $p=0.98$. When the present-day GC velocity dispersions are used, the observed IMBH masses are undeniably inconsistent with the galactic scaling relation, at $p=1.0\times10^{-6}$. Like the $M_{\rm BH}$--$M_{\rm sph}$ relation, the $M_{\rm BH}$--$\sigma$ relation extends to the GC regime when accounting for the dynamical evolution of GCs.

\section{Discussion} \label{sec:disc}
\subsection{Observational and model uncertainties}
We use a simple, {semi-analytic} model for the dynamical evolution of GCs, which has been used previously to explain several other features of GC populations, to relate observed IMBH masses to the initial masses and velocity dispersions of their host GCs. We find that these systems were initially consistent with the two main galactic scaling relations for SMBH masses ($M_{\rm BH}$--$M_{\rm sph}$ and $M_{\rm BH}$--$\sigma$), and moved off these relations due to dynamical evolution.

Despite an increasing number of 2--3$\sigma$ detections, the existence of IMBHs is still debated \citep[e.g.][]{strader12}. {Hence, we cannot rule out the possibility that} the IMBH detections are spurious, in which case it is remarkable that accounting for GC evolution makes them align with the scaling relations without fitting any free parameters. Given the uncertainties of the IMBH mass measurements, the probability that the IMBH detections are statistical noise and still exhibit the observed scatter around the relations is $p\leq3\times10^{-4}$, indicating that only systematic errors in the detection method could affect our findings. A potential issue is that the required disruption parameter $t_0$ is lower by a factor of 5--10 than expected for the current, Galactic environment of GCs. However, several earlier, independent observations have required the same shorter-than-current disruption time-scale, and Figures~\ref{fig:mbhmhost} and~\ref{fig:mbhsigma} add two more independent tests. Above all, a short disruption time-scale need not be a fundamental problem -- it is likely that GCs underwent most of their mass loss before entering their present-day environment \citep{elmegreen10,kruijssen12c}.

\subsection{A universal mechanism for black hole growth?}
While we refrain from deriving any definitive conclusions regarding massive black hole growth from our results, the agreement between IMBH and SMBH scaling relations is both intriguing and suggestive. It requires (or predicts) that while galactic SMBHs are still growing, most of the IMBH growth occurred during a rapid initial phase, when their host GCs were much denser and potentially still gas-rich. If the SMBH scaling relations reflect some imprint of the black hole formation process \citep[{and even if they simply arise from the central limit theorem, e.g.}][]{peng07}, then our results also imply that this growth process may very well be universal over seven decades in black hole mass. Universality is not necessarily contradicted by current deviations from the SMBH scaling relations -- it is important to realise that the black hole scaling relations are evolving, and tidal effects should also lead to differences between the scaling relations of brightest cluster galaxies and satellites \citep[as is indeed observed, see][]{mcgee13}.

\section*{Acknowledgments}
We are grateful to an anonymous referee for a timely report, to Anja Feldmeier, Markus Kissler-Patig, Nadine Neumayer and Tim de Zeeuw for helpful comments, and to Mark Gieles, Thorsten Naab, and Laura Sales for stimulating discussions.

\bibliographystyle{mn2e}
\bibliography{mybib}

\appendix
\section{The cluster radius evolution} \label{sec:radius}
{In this Appendix, we obtain a rough estimate for the typical radius evolution experienced by the GCs in our sample, which depends on (1) the time since accretion into the Galactic halo, and (2) the subsequent disruption time-scale. In the radius evolution model of \citet{gieles11b}, a GC first expands self-similarly to fill its Roche-lobe, after which the expansion is halted and the radius decreases as a function of time, with a GC density equal to the tidal density (i.e.~$R\propto M^{1/3}$). According to \citet{gieles11b}, Galactic GCs are in the expansion-dominated regime for $M\ga10^5~\msun(4~{\rm kpc}/R_{\rm gc})$, with $R_{\rm gc}$ the galactocentric radius. This condition is satisfied for all GCs in the \citet{luetzgendorf13} sample. These GCs have been expanding since they were accreted into the Galactic halo -- their evolution prior to that moment is unknown, and we approximate the initial GC radius by the radius at the time of accretion. \citet{rocha12} show that the median accretion time of the current Galactic satellite dwarf galaxies is $\hat{\tau}_{\rm acc}\sim6.5~{\rm Gyr}$ ago (the hat indicates a median value), but this number is naturally biased to dwarfs that have not yet merged with the Milky Way. The galaxy formation simulations of \citet{sales07} suggest that the median accretion time of all dwarf galaxies precedes that of the survivors by $\sim1.5~{\rm Gyr}$, and hence $\hat{\tau}_{\rm acc,all}\sim8~{\rm Gyr}$. Based on the current orbital parameters of our GC sample \citep{dinescu99}, we find that during these $8~{\rm Gyr}$, the median disruption parameter of the sample is $\hat{t}_0\sim7.3~{\rm Myr}$ \citep[using equation~7 of][]{kruijssen09}.

The model of \citet{gieles11b} allows us to derive the radius at the time of GC accretion as a function of the mass loss since that time and the present-day GC radius. We use the current median mass of the sample, i.e.~$\hat{M}=7.1\times10^5~\msun$, which for $\hat{t}_0=7.3~{\rm Myr}$ implies $\hat{M}_{\rm acc}=8.3\times10^5~\msun$ at the time of accretion into the Galactic halo. The projected half-light radii $R_{\rm hl,2D}$ are taken from \citet[2010~edition]{harris96}. The difference between projected and deprojected, 3D half-mass radii in the simulations of \citet{hurley07} is about a factor of two, and the possible effect of mass segregation on the conversion from a half-light to a half-mass radius is minor, especially at old ages \citep{gaburov08}. We therefore adopt a median half-mass radius $\hat{R}=2\hat{R}_{\rm hl,2D}=5.5~{\rm pc}$. With the formalism of \citet{gieles11b} we derive $\hat{R}(\hat{t}_{\rm acc})=4.4~{\rm pc}$, implying a median expansion factor since accretion of $\hat{R}/\hat{R}(\hat{t}_{\rm acc})=1.3$. As stated before, the unknown evolution of the GCs prior to the accretion into the Galactic halo forces us to use $\hat{R}(\hat{t}_{\rm acc})$ as a proxy for the initial systems. We combine $\hat{R}/\hat{R}(\hat{t}_{\rm acc})$ with the total mass loss $M/M_{\rm init}$ in equation~(\ref{eq:sigmai}) to show the effect of expansion on the initial velocity dispersions.}


\label{lastpage}

\end{document}